\documentstyle[aps,twocolumn]{revtex}

\begin{document}

\title{Laser cooling of single trapped atoms to the ground state:\\
 a dark state in position space}

\author{G. Morigi, J.I. Cirac, K. Ellinger and P. Zoller}

\address{Institut f\"ur Theoretische Physik, Universit\"at Innsbruck, \\
A--6020 Innsbruck, AUSTRIA}

\maketitle

\begin{abstract}
We propose a scheme that allows to laser cool trapped atoms to the
ground state of a one--dimensional confining potential. The scheme is based on 
the creation of a dark state by designing the laser profile, so that
the hottest atoms are coherently pumped to another
internal level, and then repumped back. The scheme works beyond the
Lamb--Dicke limit. We present results of a full quantum treatment for a  
one-dimensional model.
\end{abstract}


\date{13 October 1997}

\narrowtext

\section{Introduction}

The ultimate goal of laser cooling of neutral atoms stored in a trapping
potential is the cooling to the vibrational ground
state.
This is of interest in the context of observing effects related to
quantum statistical properties of atoms~\cite{BED} and object of
present experimental investigations~\cite{Salomon,Chu}. The hope is 
that these efforts might ultimately open a second route  to 
Bose--Einstein condensation of dilute gases~\cite{BEC,BAR}.
  
Laser cooling to the ground state of a trap has been achieved
experimentally for single trapped ions, using sideband cooling
techniques~\cite{Sideband} in the Lamb--Dicke
limit~\cite{Review1,Review2}. Sideband cooling is based on the selective
laser excitation of the low--frequency sideband in a harmonic trap, leading
 to optical pumping into the vibrational ground state. The
requirements are that the motional sidebands are spectroscopically
resolved (which is achieved in the strong confinement limit), and that
the spatial dimension of the ground state $a_0$ is much smaller the the
wavelength of the cooling laser $\lambda$ (in order to avoid the heating
produced by spontaneous emission). This latter condition is represented
by the relation $\eta\ll 1$, where $\eta$ is the Lamb--Dicke parameter
$\eta =2\pi a_0/\lambda$. In one of our recent
publications~\cite{Ignacio}, we have discussed extensions of sideband
cooling beyond the Lamb--Dicke regime. Over the last few years other
laser cooling techniques have been developed, which achieve ground state
cooling in trapping potential and which were originally developed for
free atoms~\cite{Salomon,Chu,Pellizz}. In particular, for free atoms
they have allowed to achieve temperatures below the recoil limit
$E<E_R$, corresponding to an atomic de Broglie wavelength larger than
the wavelength of the light $\lambda $. This has been obtained
experimentally, by optical pumping into a velocity selective dark
state~\cite{Dark}, or by cooling with a sequence of shaped Raman pulses
where the frequency spectrum of the light is tailored so that atoms with
near zero velocity are no longer excited~\cite{Raman}. 
These
subrecoil cooling techniques as well as sideband cooling are all
versions of ''dark state cooling''. The fundamental idea is the
decoupling of a quantum state from radiation, and the accumulation of
atoms in this state by spontaneous emission. The dark state condition
can be created with various mechanisms; in sideband cooling, the trap
ground state is dark because it is off--resonant from any other state,
since the laser is tuned to the red motional sideband. The free particle  
subrecoil cooling
schemes, instead, use dark states in momentum space. 

In the present paper we will discuss a cooling scheme which is based on
the creation of a {\em dark state in position space} with the help of an
appropriate spatial profile of the cooling laser, so that this state
(typically the ground state of the trap) is not or only weakly excited
by the laser \cite{related}. This condition can be realized, for example, using the
laser in the doughnut mode~\cite{Doughnut} with the axis aligned with
one of the trap axis. In particular, this cooling mechanism works for
Lamb--Dicke parameters $\eta\ge 1$, and outside of the strong
confinement regime (i.e. when the sidebands are not spectroscopically
resolved). We will show that the cooling scheme is quite efficient,
allowing to cool a significant fraction of atoms to the ground state of a trapping
potential. The model we will consider is one--dimensional, but the scheme is readily
extended to two and three-dimensional situations.

The paper is organized as follows. In Sec. IIA, we describe
qualitatively the cooling scheme. In Sec. IIB we develop the Master
Equation describing the system dynamics, and in IIC we discuss the
existence of a stationary solution. In Sec. III we present the numerical
results, obtained by taking a laser intensity profile corresponding to a
doughnut mode with: {\it (i)} atoms trapped by an harmonic oscillator,
and {\it (ii)} atoms trapped by another confining potential. In this last
case, the efficiency of the cooling mechanism is improved.

\section{The model}

\subsection{The laser cooling scheme}

We consider laser cooling in a three-level $\Lambda $-system, as illustrated
in Fig.~1. We denote by $|g\rangle $ and $|e\rangle $ atomic ground
(or metastable) states which are connected to the excited state $|r\rangle 
$ by dipole transitions. The atoms are confined in a one--dimensional 
harmonic potential with
oscillation frequency $\nu $. The quantized trap levels will be denoted by $
|n\rangle $.
 
The laser cooling scheme consists of the repetition of a sequence of
three pulses. In a {\em first step}, atoms in $|g\rangle $
are coherently transferred to the state $|e\rangle $ by an off--resonant
Raman pulse (Fig.~1a). The purpose of this pulse is to excite
atoms which are not in the ground state of the trapping potential. This
is achieved by assuming a spatial distribution of the laser which has
essentially zero intensity in a the region where the ground state
wavefunction $|g,0\rangle $ takes on appreciable values. On the other
hand, the spatial distribution of atoms in excited trap states overlaps
with the laser light so that these atoms will be transferred to the
state $|e\rangle $. The {\it second step} consists of laser cooling of
atoms in state $|e\rangle $, thermalizing the atomic center of mass
distribution to a few recoil energies$\ E_R$ (Fig.~1b). This is
achieved by coupling $|e\rangle $ to other internal atomic levels
employing one of the standard laser cooling schemes (e.g. polarization
gradient cooling)~\cite{Dalibard}. This is used to provide a mechanism
for confinement of atoms in a finite range of trapping levels,. In a{\em
\ third step} atoms are optically pumped into $|g\rangle $ using a laser
tuned on resonance to the transition $|e\rangle \to |r\rangle $ followed
by spontaneous transition to $|g\rangle $ (Fig~1c). We assume
that the branching ratio between the decay channels $|r\rangle \to
|g\rangle $ and $|r\rangle \to |e\rangle $ is large, so that we can
safely neglect the latter decay channel. Atoms decaying from $|r\rangle
$ to the ground state $|g\rangle $ will be distributed over a range of
trap levels of the order of a recoil energy, including the ground state
$|g,0\rangle .$ Repetition of this cooling cycle will accumulate atoms
in the spatial dark state $|g,0\rangle $.

\subsection{Master Equation}

We give here the master equation describing the evolution of the system
during the sequence of pulses. Let $\rho $ be the density matrix describing
the system of atoms interacting with the laser field. The master equation
can be written in the form~\cite{Gardiner} 
\begin{equation}
\frac d{dt}\rho (t)=L\rho (t),
\end{equation}
where $L$ is a linear superoperator whose form depends on the particular
cooling step. The formal solution is $\rho (t)=e^{L(t-t_0)}\rho (t_0)$.
The scheme consists of
three steps, and the density matrix at the end of a sequence of pulses will
be 
\begin{equation}
\rho (t_3)= e^{L_3(t_3-t_2)} e^{L_2(t_2-t_1)} e^{L_1(t_1-t_0)} \rho(t_0),
\end{equation}
where $L_i$ is the linear superoperator corresponding to
the $i$-th pulse of the sequence, and $\rho (t_0)$ is the density matrix at
the beginning of the sequence of pulses. We derive now the explicit form of
the evolution operators.

{\em First step:}{\bf \ }we assume that the evolution is coherent. We
eliminate the excited state $|r\rangle $ in perturbation theory and reduce
the dynamics to an effective two-level system $|g\rangle ,$ $|e\rangle $.
Assuming that the lasers are tuned to exact Raman resonance and
working in the frame rotating at the laser frequency, we write 

\begin{equation}
\rho (t_1)=\text{e}^{-iH(t_1-t_0)}\rho (t_0)\text{e}^{iH(t_1-t_0)},
\label{Raman_evol}
\end{equation}
where $H$ is the Hamiltonian describing the evolution during the pulse, 

\begin{equation}
H=H_0+H_{\text{I}},
\end{equation}
with 

\begin{equation}
H_0=\hbar \nu \hat{a}^{\dagger }\hat{a}  \label{H_0},
\end{equation}
and 

\begin{equation}
H_{\text{I}}=
\frac{\hbar \Omega (\hat{x})}2(\hat{\sigma}^{\dagger }e^{ik\hat{x}}+\hat{%
\sigma}e^{-ik\hat{x}}).  \label{Hint}
\end{equation}
Here, $\hat{a}$ and $\hat{a}^\dagger$ are creation and annihilation 
operators for the
harmonic oscillator, $\nu $ is the trap frequency, $\hat{\sigma}^{\dagger
}=|e\rangle \langle g|$ and $\hat{\sigma}=|g\rangle \langle e|$ are the
dipole raising and lowering operators, respectively, $\Omega (\hat{x})$ is
the Rabi frequency and $k$ the wavevector of the two--photon transition
along the (cooling) $x$ axis. 
Denoting by $\Delta t=t_1-t_0$ the time duration of this first pulse, we
will assume that $\nu \Delta t\ll 1$, so that 
we can safely neglect $H_0$ in~(\ref{Raman_evol})

\begin{equation}
\rho (t_1)\approx \text{e}^{-iH_{\text{I}}\Delta t}\rho (t_0)
\text{e}^{iH_{\text{I}}\Delta t}.
\label{Raman_evol1}
\end{equation}

{\em Second step:} Laser cooling of atoms in state $|e\rangle $ via an
auxiliar level [see Fig.~1b] destroys the
coherences between $|e\rangle$ and $|g\rangle $, while atoms left in 
$|g\rangle $ evolve according to~(\ref{H_0}). We do not specify the cooling
mechanism, assuming only that at the end of the pulse the atoms in 
$|e\rangle $ are thermalized, and described by the density matrix 

\begin{equation}
|e\rangle \langle e|\otimes \rho^{\text{th}}=|e\rangle \langle e|\otimes 
\frac{\sum_n \mbox{e}^{-n/N }|n\rangle \langle n|}{\sum_n \mbox{e}^{-n/N }},  \label{therm}
\end{equation}
where $N $ is related to the average number of vibrational quanta after
this laser cooling process. Using the properties of $\hat{\sigma}^{\dagger }$, 
$\hat{\sigma}$ in~(\ref{Hint}), we have 

\begin{eqnarray}
\rho (t_2) 
&=      &\text{e}^{-iH_0 T_{\text{sep}}}|g\rangle \langle g|\cos [\Omega (%
\hat{x})\Delta t/2]\rho (t_0) \\
&\times &\cos \left[ \Omega (\hat{x})\Delta t/2\right] |g\rangle \langle g|%
\text{e}^{iH_0 T_{\text{sep}}}+\zeta |e\rangle \rho ^{\text{th}}
\langle e|,  \nonumber
\end{eqnarray}
where $T_{\text{sep}}=t_2-t_1$ is the duration of this step, and $\zeta $ is 
the probability of occupation of $|e\rangle $, i.e. 

\begin{eqnarray}
\zeta &=&\text{trace}\{|e\rangle \langle e|\rho (t_1)\} \\
&=&1-\text{trace}\{|g\rangle \langle g|\cos \left[ \Omega (\hat{x})\Delta
t/2\right] \rho (t_0)\cos \left[ \Omega (\hat{x})\Delta t/2\right] \}. 
\nonumber
\end{eqnarray}
{\em Third step:} the thermalized atoms in $|e\rangle $ are optically pumped
into $|g\rangle $. The linear superoperator describing the master equation is 

\begin{eqnarray}
L_3\rho &= &-i[H_0,\rho ]-\frac{\Gamma}{2}\left(\rho
\hat{\sigma}^{\dagger } \hat{\sigma} + \hat{\sigma}^{\dagger }
\hat{\sigma} \rho \right) \\ \nonumber
        &+ &\Gamma \int_{-1}^1duN(u)
\hat{\sigma}
e^{ik_p(1+u)\hat{x}}\rho^{\text{th}}
e^{-ik_p(1+u)\hat{x}}\hat{\sigma}^{\dagger }.
\nonumber
\end{eqnarray}
with $\Gamma$ the optical pumping rate, $N(u)$ the angular distribution
of the emitted photons, and $k_p$ wavevector of the pumping
laser that propagates along $x$, $k_p=2\pi /\lambda =\eta/a_0$. 
Assuming $\nu ^{-1}\gg t_3-t_2$,
we can neglect the 
free evolution during this pulse. Furthermore, for $t_3-t_2\gg \Gamma
^{-1}$ all the atoms in $|e\rangle $ are pumped in $|g\rangle $ during the
pulse. Taking that into account, we find: 

\begin{eqnarray}
\label{equa_semi}
\rho(t_3)
&=      &e^{-iH_0 T_{\text{sep}}}|g\rangle \langle g|\cos [\Omega (\hat{x}%
)\Delta t/2]\rho (t_0) \\ \nonumber
&\times &\cos [\Omega (\hat{x})\Delta t/2]|g\rangle \langle
g|e^{iH_0 T_{\text{sep}}}  \\ \nonumber
&+      &\zeta \int_{-1}^1duN(u)e^{ik_p(1+u)\hat{x}}|g\rangle 
\rho ^{\text{th}}\langle g|e^{-ik_p(1+u)\hat{x}}.   
\end{eqnarray}
We can now project onto $|g\rangle $, and define
${\rho }_g=\langle g|\rho |g\rangle
$. Denoting by $F=\int_{-1}^1duN(u)e^{ik_p(1+u)\hat{x}}\rho
^{\text{th}}e^{-ik_p(1+u)\hat{x}}$ the
matrix describing the feeding contribution, we finally have 

\begin{eqnarray}
\label{final_eq}
\rho _g(t_3) &=&\text{e}^{-iH_0T_{\text{sep}}}\cos [\Omega
(\hat{x})\Delta t/2]\rho_g(t_0) \\ 
&\times &\cos [\Omega (\hat{x})\Delta
t/2]\text{e}^{iH_0T_{\text{sep}}} +\zeta ~F. \nonumber  
\end{eqnarray}
This gives a mapping between the density operator at the beginning
and at the end of a given sequence of 3 pulses.
From the above expressions, we can deduce that 
for large traps (i.e. for  large values of the Lamb--Dike parameter) 
the cooling will slow down and become less efficient. The reason for that
is twofold: on one hand, the kick provided by the optical pumping
process will distribute the atoms among a wider range of states. On the
other hand, since we have assumed that the thermal distribution 
$\rho^{\text{th}}$  
has a mean energy of the order of the recoil, the mean occupation number
will be larger ($\langle n\rangle \approx E_R/\hbar\nu=\eta^2$).

\subsection{Stationary state}

Ideally, we would like to use the cooling mechanism to cool all the
atoms to the ground state (or any other pure state). Therefore,
we have to analyze the shape of $\Omega(x)$ which gives rise
to the optimal cooling. In this subsection we analyze qualitatively
the conditions for the spatial profile of the laser $\Omega(x)$ to
provide an efficient cooling. 

We then look for stationary solutions to the mapping (\ref{final_eq})
which corresponds to pure states. A necessary condition for that is 
$\zeta =0$. Then, we substitute $\zeta =0$ in~(\ref{final_eq}) and
impose the stationary regime, i.e. $\rho _g(t_3)=\rho
_g(t_0)=\rho _g$, with
$\rho _g=|f\rangle \langle f|$. We obtain that these conditions are
satisfied if the following equation allows for a solution: 

\begin{equation}
\langle x|\text{e}^{iH_0T_{\text{sep}}}|f\rangle \text{e}^{i\phi }
=\cos [\Omega (x)\Delta t/2]\langle x|f\rangle,  \label{stazio2}
\label{Giovanna}
\end{equation}
with $\phi $ arbitrary phase. Let us denote by $R$ the region where
$\langle x|f\rangle $ is different from zero. 
It can be easily shown that~(\ref{Giovanna})
implies that in the region $R$ the relation 
$|\cos [\Omega (x)\Delta t/2]|=1$ must be fulfilled. Physically,
this means that if we want to have a dark state $|f\rangle$ 
its wavefunction must be completely distributed in the spatial region 
R where there is no laser excitation (that is, where the atoms perform
complete Rabi oscillations). On the other
hand, we want to have only one dark state, and therefore other states
must have a finite excitation probability. This means that
they must be nonzero outside $R$. Therefore, for achieving cooling to the 
ground state, one should design the intensity profile of the laser in such 
a way that only in the region
where the ground state is localized, the laser accomplishes no
excitation. This condition can be only approximately fulfilled
in a harmonic trap, where the ratio among the spatial dimensions of the ground
state and the first excited state is $1/\sqrt{2}$. In this case, we 
expect to not 
be able to cool all the atoms into the ground state. Cooling into the ground 
state will be possible in the case of atoms trapped by a confining potential,
which better localizes the ground state respect to the first excited state.

The qualitative argument given above does not depend on the separation 
time $T_{\rm sep}$
in the second pulse. In reality, the selection of this time can play an important
role, especially in the case of a harmonic potential. To see that, let us
consider the particular case
$\nu T_{\text{sep}} = 2l\pi$, with $l$ integer. Then, according to
(\ref{final_eq}) the positions $x$ in the region $R$ in which
$|\cos [\Omega (x)\Delta t/2]|\neq 1$ will be emptied, whereas the atoms will
be accumulated in the regions where 
$|\cos [\Omega (x)\Delta t/2]|= 1$. In fact, 
from~(\ref{Giovanna}) we see that the state $|f\rangle$ is a stable solution 
if the conditions $\cos(\Omega(x)\Delta t/2)|f\rangle = 
\text{exp}(i\phi_1)|f\rangle $ and if 
$\text{exp}(-iH_0T_{\text{sep}})|f\rangle 
= \text{exp}(i\phi_2)|f\rangle$ are fulfilled, with $\phi_1+\phi_2=\phi$.
For $\nu T_{\text{sep}} = 2l\pi$ this means that each
wavepacket completely distributed inside $R$ at $t_0$ is solution (i.e. a dark state),
since in $T_{\text{sep}}$ it undergoes a full oscillation and therefore
recovers the original form. In that case, laser cooling will not
be possible since there will exist many wavepackets that will remain dark.
In the case $\nu T_{\text{sep}} = (2l+1)\pi$, one can consider two subspaces
$S_1=\{|0\rangle,|2\rangle,|4\rangle,...\}$ and  
$S_2=\{|1\rangle,|3\rangle,|5\rangle,...\}$, in which the evolution during
the second pulse is periodic. That is, if we can form a wavepacket as
a linear combination of the states of $S_1$ or of $S_2$ which is spatially
distributed in $R$ at $t_0$, this wavepacket will be a dark state. Hence,
laser cooling will not be possible either. The argument can be repeated
now for times fulfilling $\nu T_{\text{sep}} = n\pi/m$, and subspaces 
can be constructed so that wavepackets belonging to them will be
dark states, and therefore laser cooling will not be possible.
We can overcome this problem very easily by choosing {\it random separation
times} $T_{\rm sep}$ so that the only wavepacket that is a dark state
is precisely the ground state.

\section{Results of calculations}

In the following, we show results obtained assuming a doughnut mode as
spatial distribution of the Raman pulse. The reason for this choice
is that it provides a flat profile in a finite region which, as we
have shown above, is required to obtain a suitable cooling.
In particular, we discuss 
the parameter regime of the laser for which the cooling efficiency is
optimal. Finally, we explore the role of
$T_{\text{sep}}$ in the cooling dynamics. We analyze first
the case of a harmonic 
trapping potential and then another potential which is best suited for laser
cooling using the present scheme.

We take as spatial distribution for the Raman pulse laser the following
class of functions: 
\begin{equation}
f(x)=\left(\frac{x}{a_0}\right)^{2n}
\exp \left( -\frac 12\left( \frac{x/a_0}\alpha \right) ^2\right) .
\label{Doughnut}
\end{equation}
Here $\alpha $ determines the spatial width of the pulse in units of the
harmonic oscillator ground state width $a_0$, and the exponent $n$ is
an integer 
number, which denotes the order of the doughnut mode and thus determines the
flatness around $x=0$. In the following, we explore the dependence of the
cooling efficiency on the parameters $n$ and $\alpha $, for atoms whose
center of mass is trapped by a harmonic potential with frequency $\nu $.
We take the Rabi frequency $\Omega(\hat{x}) \propto f(\hat{x})$.
Furthermore, the projection along the cooling axis ($x$) of the two-photon 
wavevector
appearing in~(\ref{Hint}) is $k=2\pi /\lambda $, where $\lambda $ is the
wavelength of the transition $|r\rangle \to |e\rangle $, since the doughnut
mode photons propagate in a direction orthogonal to the $x$ axis. 

In Fig.~2a and Fig.~2b we show the population
$P_g^0=\langle g,0|\rho_g|g,0\rangle $ after 1500 (curve with $\circ$) and 2500
(curve with $\times $) sequences of pulses as a function 
of the width $\alpha $ and of the exponent $2n$, respectively. We have obtained
these results using Eq.~(\ref{final_eq}), with 
a Lamb--Dicke parameter $\eta =2\pi a_0/\lambda =5$, and with a 
thermalized distribution $\rho ^{\text{th}}$  in
Eq. (\ref{therm}) with $N =\eta ^2=25$. From
Fig.~2a the existence of an 
optimal width $\alpha $ for a given exponent (in the present case: $2n=4$)
is evident. The optimal $\alpha $ depends on $n$, since it must fulfill the
requirements of Eq.~(\ref{stazio2}). In Fig.~2b we plot $P_g^0$ as a
function of $2n$, where the values of the width have been optimized. We see
that for $n\ge 4$, around $80\%$ of the atoms are cooled to the ground state.
Figure~3a shows the population $P_g^n=\langle n|\rho
_g|n\rangle $ as a function of the vibrational number $n$ after 
2500 sequences of pulses
for $n=4$ and $\alpha =4$, and Fig.~3b shows $P_g^0$ as a
function of the number of pulses. In the inset of this figure the final spatial
distribution is shown (solid line), while the spatial distribution $\cos
[\Omega (x)\Delta t/2]$ is plotted as a dashed line for comparison. In all
the cases, the final density matrix is not a pure
state but a mixture, diagonal in the number states basis. According to
the discussion in the previous Section, we have chosen
random separation times. We have also checked that by fixing the separation
times $T_{\rm sep}$, the cooling efficiency decreases significatively.

The efficiency of the cooling scheme in populating the ground state depends
on the difference between the reexcitation rate of the vibrational
ground state and the ones of the first few excited states. To the
extent that one can design a 
confining potential so that the ground state is much more localized in space
than the other eigenstates, the cooling efficiency will improve
correspondingly. A confining potential satisfying the above requirement
could be experimentally realized, for example, by adding to an harmonic trap
an additional optical potential. In the following we will
assume a potential 
\begin{equation}
V(x)=\frac{1}{2}2m\nu ^2\left[ x^2+\epsilon \frac{x^2}{1+gx^2}\right],
\label{perturb}
\end{equation}
where $\epsilon $ and $g$ characterize the shape of the potential
near the center. An example
of how the potential ~(\ref{perturb}) looks like is shown in Fig.~4a.
Figure 4b shows the spatial widths of the first
twelve eigenstates for $g=2000/a_0^2$ and $\epsilon =19\times 10^{5}$ 
(curve with $\circ$), compared with the
corresponding widths of the harmonic oscillator eigenstates with the same 
$\nu $ (curve with $\times $). For this choice of the parameters 
$\epsilon $ and $g$,
the eigenvectors of the potential tend to the number states of the harmonic
oscillator for $n\ge 1$. In the present example, the ground state has spatial
width $a_0/5$, and it is thus well localized respect to the other
eigenstates.
The laser cooling scheme works very efficiently in populating the ground
state, as it may be seen from Fig~5: here the population of the
ground state in the trapping potential $V(x)$ is plotted as
a function of the number of pulses (solid line) for a doughnut mode with $\alpha
=8$ and $2n=2$ [c.f. Eq.~(\ref{Doughnut})]. The Lamb--Dicke parameter for the
eigenstates $n\ge 1$ and for the internal excited state $|e\rangle $, is 
$\eta =5$, while the ground state has spatial width corresponding to an
effective Lamb--Dicke parameter $\eta ^{\prime }=\eta /5=1$. We can see that
during the cooling almost all atoms are accumulated in the ground state of $V(x)$. 
The dashed line in Fig.~5 shows for reference the evolution
of the population of the ground state of a pure harmonic oscillator
potential with $\eta =1$. In this second case, we have used a doughnut mode
with $2n=2$ and $\alpha =10$. Note that the sample with $\eta =1$ is colder
at the beginning, since its initial distribution is thermalized with $\langle
n\rangle\approx 4$.

The result obtained here is rather general and can be summarized as follows: 
through the creation of a more
localized ground state, obtained by a suitable perturbation of an
harmonic potential of a given $\eta$, the present cooling mechanism 
allows to cool all the atoms to the ground state, even for the Raman 
pulse intensity distribution in a low order doughnut mode.

\section{Conclusions}

In this paper we have studied laser cooling of trapped atoms to a
single quantum state, based on creating a dark state in position space
by reexciting the atoms with cooling laser which has no spatial overlap
with the trap ground state. The scheme can be applied in the limit of
weak confinement ($\Gamma> \nu$) and outside the Lamb--Dicke regime. Assuming a doughnut
mode as a model of the spatial profile of the laser, our numerical
calculations have shown that more than $80\%$ of the atoms
can be cooled to the ground state of a harmonic oscillator
potential. Furthermore, this number can be increased significantly for
trapping potentials where the ground state is much better localized than
the first few excited vibrational states. 
The randomization of the separation time in the harmonic oscillator case
is fundamental in order to achieve laser cooling to the ground state. 

The model presented here is one dimensional, but it can be easily
extended to two dimensions. In that case, one could use the plane
orthogonal to the doughnut mode axis obtaining similar results as in the
one dimensional case. The reason for that is that the heating processes
(due to photon scattering during repumping) that we have considered in
the one dimensional case will be similar to the two dimensional one.
Experimentally, the one--dimensional and
two--dimensional schemes correspond to the employement of a single
doughnut mode, whose axis coincides with the one and two--dimensional
trap axes respectively, and which dark region is shaped into the size of
the one--dimensional and two--dimensional trap ground state
respectively. The extension to three dimensions could be achieved
crossing two doughnut modes, which propagate in directions
that are orthogonal one to the other, and whose axes cross at the
center of a three dimensional trap. In this way a three dimensional
dark region would be created, that could be shaped into the size of the
trap ground state.
The scheme may be useful for achieving Bose--Einstein
condensation with laser cooling, and more in general in preparing 
with laser cooling a macroscopic occupation of a state of motion.

Acknowledgment: G.M. is grateful to J.~Eschner for many stimulating 
discussions.
This work was supported by the Austrian
Fond zur F\"orderung der wissenschaftlichen Forschung and the TMR network 
ERBFMRX-CT96-0002.

\begin{figure}
%\begin{center}
%\epsfig{file=fig1.ps,width=\linewidth}
\begin{caption}
{
Internal configuration of the atom, and scheme of the sequence of pulses.
Internal configuration: $|g\rangle$, $|e\rangle$ stable or metastable states,
$|r\rangle$ excited state, with 
$|g\rangle\to |r\rangle$, $|e\rangle\to |r\rangle$ dipole transitions. 
Scheme of the sequence: (a) coherent Raman pulse, which corresponds 
to a doughnut mode on $|g\rangle\to |r\rangle$; (b) cooling 
to the recoil on the level $|e\rangle$, obtained coupling
$|e\rangle$ to a fourth internal level. (c) optical pumping into $|g\rangle$, 
obtained tuning a laser on resonance on $|e\rangle\to|r\rangle$. We 
assume that the branching ratio between $|r\rangle\to |g\rangle$ and 
$|r\rangle\to |e\rangle$ is very large, 
so that the decay along $|r\rangle\to |e\rangle$ can be safely neglected.
}
\end{caption}
%\end{center}
\end{figure}

\begin{figure}
%\begin{center}
%\epsfig{file=fig2.ps,width=\linewidth}
\begin{caption}
{
Plots of the efficiency after 1500 (curve with $\circ$) 
and after 2500 (curve with $\times$) sequences of pulses, in function of
(a) the width $\alpha$, for $2n=4$ and (b) the exponents $2n=2,4,6,8$, with
optimized widths, respectively $\alpha=30,4.2,3,2.2$. In all cases 
$\mbox{max}(\Omega(x)\Delta t/2)=0.6\pi$. $\nu T_{\text{sep}}$ is random, 
taken from a flat distribution that varies in the interval 
$\left[0.1,1.1\right] $.
}
\end{caption}
%\end{center}
\end{figure}

\begin{figure}
%\begin{center}
%\epsfig{file=fig3.ps,width=\linewidth}
\begin{caption}
{
(a) Population in function of the vibrational number state 
after 2500 sequences of pulses, for a harmonic oscillator with $\eta=5$ 
and Raman laser corresponding to a doughnut mode with $2n=4$,
$\alpha=4$ and $\mbox{max}(\Omega(x)\Delta t/2)=0.6\pi$. 
Inlay: spatial distribution of the sample after 2500 pulses (solid line) 
and $\cos(\Omega(x)\Delta t/2)$ (dashed line) plotted for comparison. 
(b): Ground state of the trap population in function of the number of pulses, 
under the same condition as above. The initial distribution and the excited 
state distribution after the recoil cooling are described by 
$\rho^{\text{th}}$ with $N=25$. $\nu T_{\text{sep}}$ is random, 
taken from a flat distribution that varies in the interval 
$\left[0.1,1.1\right] $.
}
\end{caption}
%\end{center}
\end{figure}

\begin{figure}
%\begin{center}
%\epsfig{file=fig4.ps,width=\linewidth}
\begin{caption}
{
(a): plot of $V(x)$ for arbitrary values of $\epsilon$ and $g$. (b): 
$\langle \Delta x^2\rangle^{1/2}$
in function of the first 12 eigenstates of $V(x)$ (curve with $\circ$) 
and of the first 12 ones of the Harmonic oscillator potential (
curve with $\times$), for $\epsilon=19\times 10^5$ and 
$g=2000/a_0^2$.  
}
\end{caption}
%\end{center}
\end{figure}

\begin{figure}
%\begin{center}
%\epsfig{file=fig5.ps,width=\linewidth}
\begin{caption}
{
Ground state occupation probability versus the pulse number. Solid
line: ground state of atoms trapped by $V(x)$, with $\eta=5$ and 
$\eta^{\prime}=1$, $\rho^{\text{th}}$ with $N=25$ and Raman laser 
corresponding to a doughnut mode with $2n=2$ and $\alpha=8$. 
Dashed line: ground state of an harmonic oscillator with $\eta=1$, 
$\rho^{\text{th}}$ with $N=4$ and Raman laser 
corresponding to a doughnut mode with $2n=2$ and $\alpha=10$.
In both cases $\mbox{max}(\Omega(x)\Delta t/2)=0.6\pi$. 
$\nu T_{\text{sep}}$ is random, 
taken from a flat distribution that varies in the interval 
$\left[0.1,1.1\right] $.
}
\end{caption}
%\end{center}
\end{figure}

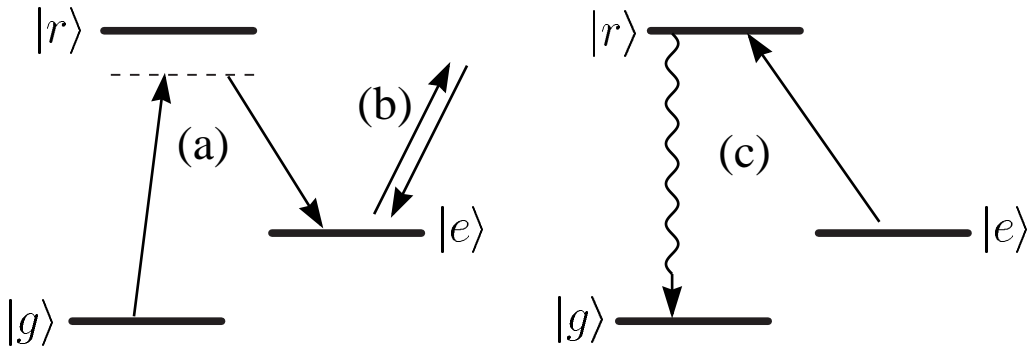
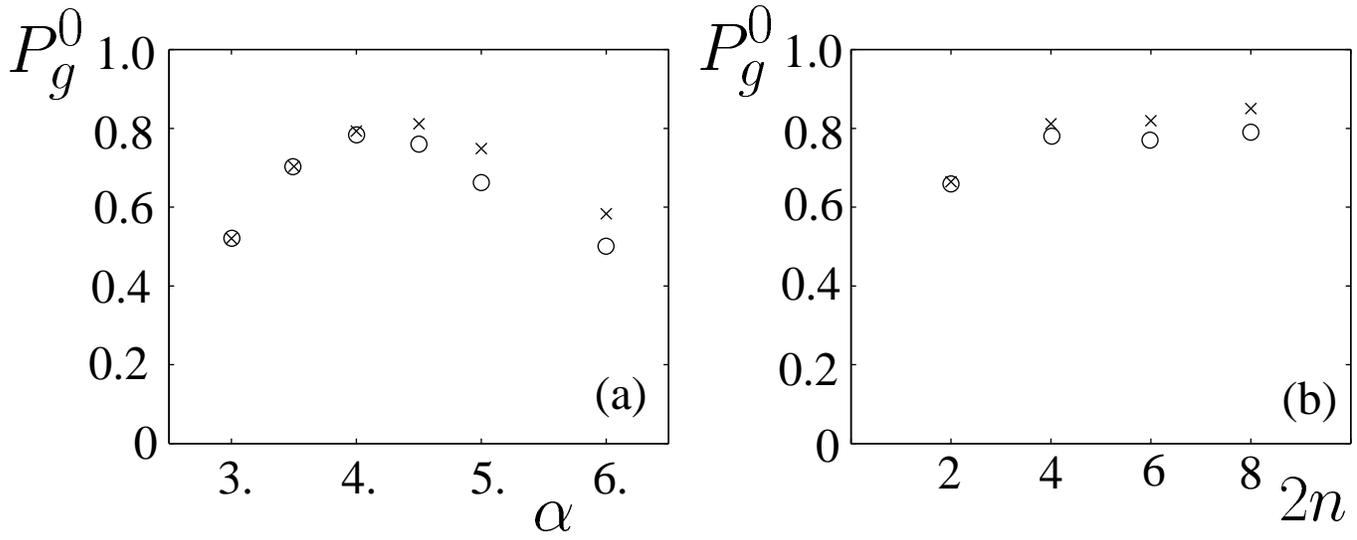
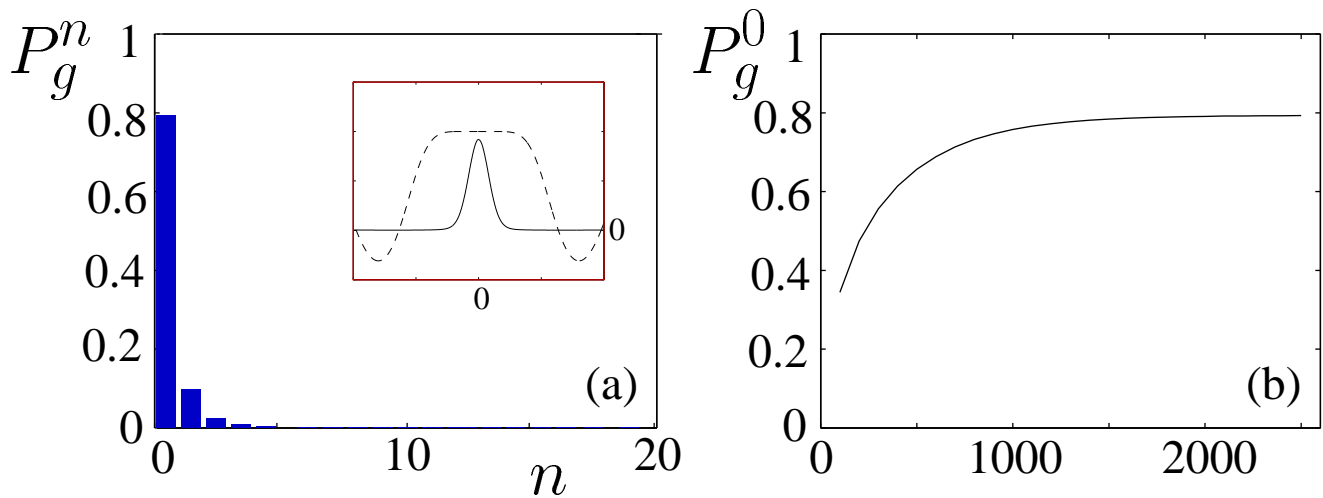
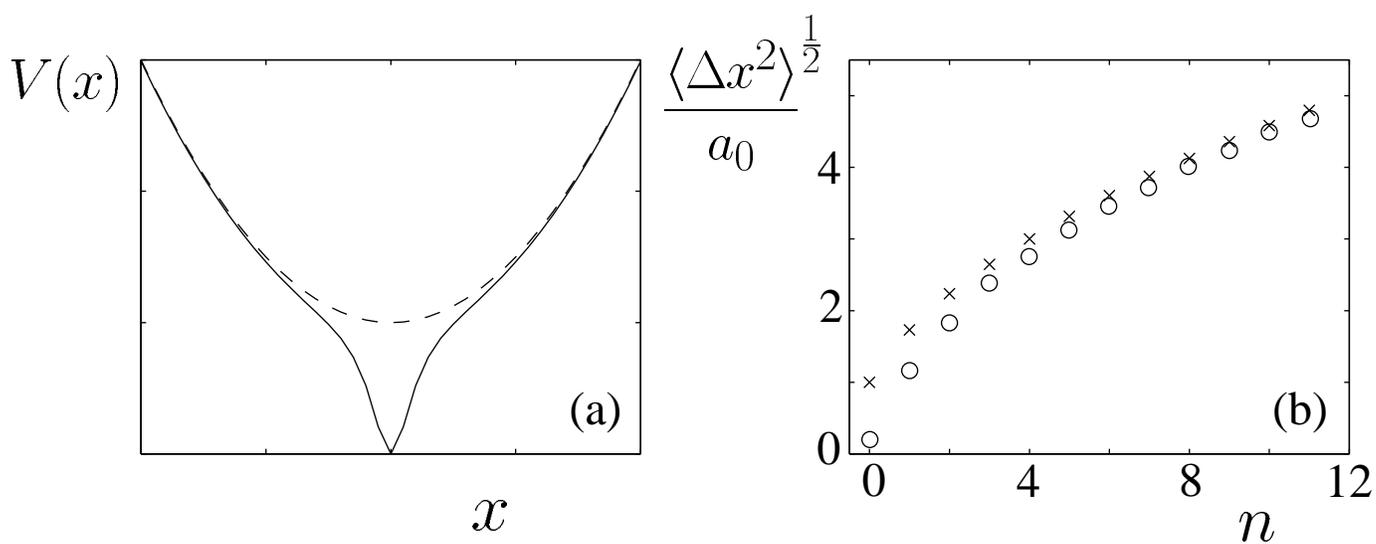
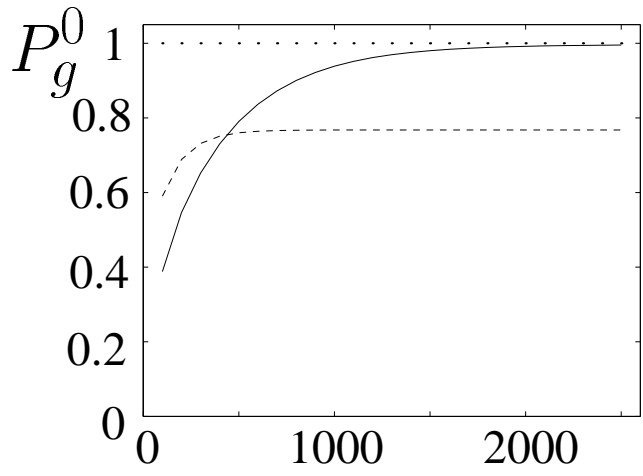
\end{document}